\documentclass[runningheads]{llncs}
\usepackage[dvipdfmx]{graphicx}
\usepackage{bm}
\usepackage{amsmath}

\begin{document}
\title{On Parallelism in Music and Language:\\ A Perspective from Symbol Emergence Systems based on Probabilistic Generative Models\thanks{This paper was written as a post-proceedings paper for the keynote speech titled ``Generative Models for Symbol Emergence based on Real-World Sensory-motor Information and Communication'' presented at the 15th International Symposium on Computer Music Multidisciplinary Research (CMMR) 2021.
This work was supported by JSPS KAKENHI Grant Numbers JP16H06569 and JP21H04904.}}
\titlerunning{On Parallelism in Music and Language from Symbol Emergence Systems}
%
\author{Tadahiro Taniguchi\inst{1}\orcidID{0000-0002-5682-2076}
}
\authorrunning{T. Taniguchi}
%
\institute{Ritsumeikan University, 1-1-1 Noji Higashi, Kusatsu, Shiga 525-8577, Japan 
\email{taniguchi@em.ci.ritsumei.ac.jp}\\
\url{http://www.em.ci.ritsumei.ac.jp/} 
}
\maketitle              
\begin{abstract}

Music and language are structurally similar. Such structural similarity is often explained by generative processes. This paper describes the recent development of probabilistic generative models (PGMs) for language learning and symbol emergence in robotics. Symbol emergence in robotics aims to develop a robot that can adapt to real-world environments and human linguistic communications and acquire language from sensorimotor information alone (i.e., in an unsupervised manner). 
This is regarded as a constructive approach to symbol emergence systems.
To this end, a series of PGMs have been developed, including those for simultaneous phoneme and word discovery, lexical acquisition, object and spatial concept formation, and the emergence of a symbol system. By extending the models, a symbol emergence system comprising a multi-agent system in which a symbol system emerges is revealed to be modeled using PGMs. In this model, symbol emergence can be regarded as collective predictive coding. This paper expands on this idea by combining the theory that ''emotion is based on the predictive coding of interoceptive signals'' and ''symbol emergence systems,'' and describes the possible hypothesis of the emergence of meaning in music.

\end{abstract}
\section{Introduction}
Symbol emergence in robotics (SER) is a constructive approach for symbol emergence systems~\cite{Taniguchi2016SER}.
Humans use symbol systems including language. To build an artificial cognitive system that can adapt to a society in which humans use symbols in an adaptive manner and understand human intelligence that can let symbol systems evolve, emerge, learn, and be used, we must understand the dynamics of symbol emergence systems in a constructive manner~\cite{Taniguchi2016SER,taniguchi2018symbol}. 
A series of studies on SER have attempted to reproduce cognitive behaviors that enable humans to acquire language and form internal representations and external symbol systems with artificial robotic and computational models~\cite{Taniguchi2016SER}. For example, researchers have developed cognitive developmental robots that perform multimodal object and place categorization and automatic phoneme and word discovery~\cite{Cangelosi2015,Ando,Araki2011,Araki2012,Nakamura2014,Akira2016,Akira2017,Hagiwara2018,Taniguchi2016,Taniguchi2016b}. Importantly, many of these are performed through unsupervised learning using probabilistic generative models (PGMs)~\cite{Bishop}.

It is noteworthy that these studies are implicitly motivated by and appear to be related to parallelism in language and music. Studies on automatic lexical acquisition by robots in which the robots form object (or place) categories and discover words from multimodal sensorimotor information and speech signals by mutually segmenting and integrating them have been inspired by the ``mutual segmentation hypothesis'' proposed in relation to studies on songbirds~\cite{OKANOYA2007271,okanoya2007neural,okanoya2017sexual}. Language models used in modeling phonemes or word sequences have been found to be naturally applied to automatic music composition. Importantly, the view analogy (or similarity) is concreted by reproducing the behaviors using computational models, that is, PGMs. 

Considering this context, it will be worth revisiting parallelism in music and language from the viewpoint of SER and symbol emergence systems. Recently, a new computational model has been proposed that models symbol emergence systems and enables computation agents to emerge symbols as a decentralized Bayesian inference~\cite{taniguchi2022emergent,Hagiwara2019,hagiwara2022multiagent}. This is based on a new type of language game, namely, the Metropolis Hastings naming game. The model and its results suggest that symbol emergence in a society can be considered as collective predictive coding. In other words, we can hypothesize that the emergence of symbols and their meanings can be modeled from the viewpoint of predictive coding~\cite{hohwy2013predictive}. What will the findings and hypotheses suggest regarding the emergence of music and the meaning of music as a symbol system? It may be worth exploring the possible hypotheses suggested by symbol emergence systems and their PGM-based models.

The research question here is ``How can the view of symbol emergence systems contribute to the discussion of the meaning of music?'' This question is crucially related to ``What is the meaning of music?'' To answer this question, I would like to present a hypothetical argument based on studies on symbol emergence in robotics, which has been conducted based on a probabilistic generative model (PGM), and a recent understanding of ``emotion.'' Recently, the idea of understanding ``emotion'' from the viewpoint of predictive coding has become prevalent~\cite{seth2013interoceptive,barrett2015interoceptive}. 
By replacing the perceptual internal representations in symbol emergence systems with emotional ones and replacing the physical interaction with the external environment, that is, the world, using the sensorimotor system with interactions with the internal environment (i.e., body) via introspective systems, I hypothetically propose parallelism in music and language from the perspective of symbol emergence systems (Figure~\ref{fig:ses}).  


The remainder of this paper is organized as follows.
Section 2 reviews a series of studies on symbol emergence in robotics and automatic music composition in computers. Through this review, we will identify a type of parallelism in music and language from the viewpoint of PGMs. Section 3 briefly introduces the view of symbol emergence systems and the concept of collective predictive coding as an account of symbol emergence. 
Section 4 presents a hypothetical view of the ''meaning of music'' from the viewpoint of symbol emergence systems. Finally, Section 5 concludes the study.

\section{Language acquisition and music composition using PGMs}

\subsection{Multimodal Concept Formation and Lexical Acquisition in Robotics}
Language acquisition by infants is closely connected to their multimodal sensory-motor information. 
A series of studies on symbol emergence in robotics have attempted to reproduce the language acquisition process using machine learning models and robots~\cite{Taniguchi2016SER}. 
By integrating multimodal information into a PGM, a computational model of language acquisition enables a robot to acquire grounded lexicons to some extent~\cite{Nakamura2007,nakamura_grounding,Nakamura2012a,Nakamura2015}.

Okanoya et al. focused on the articulate structure and grammar of songs sung by songbirds and proposed the ``mutual segmentation hypothesis'' of song phrases and song contexts~\cite{OKANOYA2007271,okanoya2007neural,okanoya2017sexual}. The hypothesis is based on the idea that segmentation of context, which is also considered as a categorization of objects and situations, and segmentation of strings, for example, speech signals, are mutually dependent and indicate that the two weakly coupled processes are the basis of human language acquisition. This hypothesis motivated me to conduct a series of studies along with collaborators.

An important step in language acquisition is word discovery and phoneme acquisition.
In artificial intelligence research, speech recognition typically means ``text to speech,'' and usually, speech signals and their transcriptions are prepared. The speech recognition systems are trained using these systems. However, language acquisition in human infants is different. Infants cannot read transcriptions, that is, written texts, before acquiring spoken language.

The generative process of a spoken utterance $\bm y$ is described as follows:

\begin{align}
 \bm{w} &= w_{1:S} \sim p(\bm{w}|\bm{z}), \label{eq:z2w}\\
 \bm{y} &= y_{1:T} \sim p(\bm{y}|\bm{w}) \label{eq:w2y}
\end{align}
where $\bm{y} = y_{1:T}$ is the acoustic feature, that is, speech signals, corresponding to the word sequence, and $z$ is a cause, that is, a state of the internal representation that generates the semiotic sign, that is, utterance.
Here, conventional speech recognition is considered an inference of $p(\bm{w}|\bm{y})$, assuming that the learning system can obtain both $\bm{w}$ and $\bm{y}$ in the training datasets, although this assumption cannot be applied to human child development.  

Unsupervised simultaneous phoneme and word acquisition was achieved by modeling the generative process of speech signals with a PGM and inferring latent variables representing phonemes and words~\cite{Taniguchi2016,Taniguchi2016b}.
Speech signals have a two-layer hierarchical structure called double articulation, which is also called the ``duality of patterning.'' This means that a speech signal is grouped into phonemes, and the phonemes are segmented into words~\cite{Chandler2002}. 
If we describe phoneme sequence $\bm l$ explicitly, the generative process shown in (\ref{eq:w2y}) becomes

\begin{equation}
    \bm{y} \sim\sum  p(\bm{y}|\bm{l}) p(\bm{l}| \bm{w}).\label{eq:w2l2y}
\end{equation}

The Bayesian double articulation analyzer is based on a nonparametric Bayesian PGM called the hierarchical Dirichlet process-hidden language model~\cite{Taniguchi2016}. The generative process simply models a double articulation structure that represents (\ref{eq:w2l2y})~\cite{Chandler2002}. Based on the PGM, a Bayesian inference procedure for phoneme and word discovery, that is, the sampling procedure of $p(\bm{w},\bm{l}|bm{y})$, was developed. 
It is known that infants use not only distributional cues, which are sound sequence information, but also prosodic cues, which are prosody information (accent and silent intervals), and co-occurrence cues, which represent co-occurrence information with other stimuli, such as multimodal sensorimotor information, for lexical acquisition~\cite{Saffran1996a}. Models that integrate these two have also been proposed~\cite{okuda2022double,taniguchi2022unsupervised,Nakamura2014,taniguchi2018unsupervised}.

Since 2015, unsupervised training of speech recognition systems has received considerable attention owing to the Zerospeech challenges~\cite{dunbar2017zero,van2020vector,tjandra2020transformer}. The performance of unsupervised speech recognition systems has been significantly improved, especially in relation to representation learning based on self-supervised learning using neural networks (e.g., ~\cite{baevski2020wav2vec}).

There have also been a series of studies on object concept (or category) formation by robots based on multimodal information. By integrating multimodal information such as visual, haptic, and auditory information using a PGM, it has been shown that robots can form object categories at various levels in an unsupervised manner~\cite{nakamura_grounding,Ando,Nakamura2015}. Similar studies have been conducted on place categories~\cite{Akira2017,Akira2016,Hagiwara2018}.It was also shown these concepts can be used for planning and active perception~\cite{Isobe-tidy-up,Akira2020,taniguchi2018multimodal}. These studies have shown that predictive coding based on PGMs can represent the formation of internal representations based on physical interactions in symbol emergence systems.

In these studies on multimodal concept formation, the generative process of multimodal sensorimotor information is described as follows.

\begin{equation}
 \{ \bm{o}_m \} \sim p(\{\bm{o}_m\}|\bm{z}) 
\end{equation}
where the generative process means that an agent, for example, a person or a robot, attempts to ``predict'' sensorimotor information. In this equation, $\bm{o}_m$ represents sensorimotor information of the $m$-th modality, and $\bm z$ represents the internal cause, that is, the perceptual state of the internal representations.
Therefore, the learning process of concepts and categories corresponds to the inference of internal representations. 

\begin{equation}
 \bm{z} \sim p(\bm{z}|\{\bm{o}_m\}) 
\end{equation}

This model is based on predictive coding in a broad sense. It is assumed that agents form concepts and categories to improve the prediction performance for multimodal sensorimotor information.

Furthermore, the PGMs proposed in these studies could be integrated. Grounded lexical acquisition from speech signals has been achieved by integrating models of unsupervised phoneme and word acquisition and multimodal category formation. Mutual learning between object categories and speech signals and between place categories and speech signals has been described\cite{taniguchi2022unsupervised,Nakamura2014,taniguchi2018unsupervised}. 
Lexical acquisition from speech signals is achievable by integrating models of unsupervised phoneme and word acquisition and multimodal category formation\footnote{Thus, integrating various predictive coding with PGMs is essential for modeling integrative human cognitive systems. As a framework for this purpose, SERKET was proposed~\cite{nakamura2018serket,taniguchi2020neuro}. Recently, the idea was extended to the whole-brain PGM, which aims to build a cognitive model covering an entire brain by combining PGMs with anatomical knowledge of brain architecture~\cite{taniguchi2022whole}. This approach is known as the whole-brain architecture approach~\cite{yamakawa2016whole}. Following this idea, the anatomical validity of the above NPB-DAA for spoken language acquisition and SLAM-based place recognition was also examined from the viewpoint of the brain~\cite{taniguchi2022hippocampal,AkiraTaniguchi2022}}.

Overall, several studies on symbol emergence in robotics have developed a wide range of learning methods assuming generative models representing
\begin{align}
 \bm{w}, \{\bm{o}_m\} &\sim  p(\bm{w}, \{\bm{o}_m\} | \bm{z}),\\
 \bm{y} &\sim p(\bm{y}|\bm{w}).
\end{align}
If we assume $\bm{z}$ is a discrete variable and iterative, i.e., mutual, inference of $\bm{z}$ and $\bm{w}$ can be regarded as mutual segmentation of strings and contexts mentioned in ``mutual segmentation hypothesis~\cite{okanoya2017sexual}.''

\subsection{Automatic Music Composition in Computers}

The similarity between music and language lies first in the fact that they are represented by a linear series of discrete sounds.
Simply put, language is a series of phonemes or letters, whereas music is a series of sounds of a certain pitch or note. Many aspects of language and music are not bound by discreteness, such as the expression of emotion through prosody in spoken language and pitch bend in musical performance. However, especially in written documents and musical scores, the use of discrete letters or a series of notes is an acceptable approximation.

In linguistics and information science, ``How do we model language?'' is the major question. However, if we simply consider strings of letters or word sequences as simple arrays of discrete ``symbols,'' it was a mathematically valid attempt to model them as stochastic processes of strings of letters or word sequences. This is a language model. This idea even goes back to Shannon's paper, which is a classic in information theory~\cite{shannon1948mathematical}.

When considering a word sequence (or string) $w_{1:S} =\{w_1, w_2, \ldots, w_S \}$, the language model is a statistical model that computes $P(w_{1:S})$. In many cases, $P(w_t|w_{1:t-1})$ is considered to generate a word sequence in time direction $t$. An approximation of $P(w_t|w_{t-n+1:t-1})$, which is censored at $n$, is called an n-gram model.
Until the mid-2010s, the n-gram model was the standard approach for modeling $w_{t-n+1:t}$.
However, since the success of deep learning in the 2010s, methods that directly approximate $P(\cdot |w_{1:t-1}) \approx f(w_{1:t-1}; \theta)$ using neural networks such as LSTM have become a new standard approach. Here, $P(\cdot)$ indicates that the probability of all random variables is output as a vector. In addition, $f$ is a function represented by the neural network, where $\theta$ is its parameter\footnote{Recently, this idea has been developed into a large-scale language model using transformers, and its generality and performance have become widely known.}.

The language model described above is a type of PGM. This represents the stochastic process by which a discrete series of words (or letters) is generated. The difference between music and language syntax has been discussed, but there is still a difference. There is a difference between music and language in that music does not have clear grammar, such as the double articulation structure and phrase structure grammar found in language. However, most language models do not explicitly consider these factors~\cite{asano2015syntax}. Therefore, the concept of the language model can be used equally well to capture the syntax latent in sound sequences in music. The language model is an important abstraction when discussing the parallelism between language and music.

In a musical performance, the actual sound $y$ is generated by a performer. If we consider the generative process of music, including composition and performance, it can be described as follows: 

\begin{align}
 \bm{w} &= w_{1:S} \sim p(\bm{w}|\bm{z}), \label{eq:z2w_}\\
 \bm{y} &= y_{1:T} \sim p(\bm{y}|\bm{w}) \label{eq:w2y_}
\end{align}
where $\bm{z}$ is the cause of music generation, for example, the emotional state with which a player and composer attempt to generate the song.  
Interestingly, these equations are identical to those in (\ref{eq:z2w}) and (\ref{eq:w2y}), respectively.
This correspondence apparently displays one parallelism between music and language.

Many statistical language models have been used to capture note sequences for automatic compositions. The bigram and trigram models of notes are not sufficient to capture the long-term dependency. Therefore, longer context-aware language models are required. A language model based on nonparametric Bayesian methods that consider a theoretically infinite number of contexts was proposed. Shirai et al. proposed a melody generation method using the variable-order Pitman–Yor language model proposed by Mochihashi et al.~\cite{mochihashi2007infinite}. Another key point for automatic composition using the PGM is that ``sampling from the posterior distribution'' can be explicitly considered.
In creative activities such as composing music, it is more important to propose a reasonably large number of candidates than to find the optimal solution.
This means that rather than formulating automatic composition as an optimization problem, it is more appropriate to consider it as sampling from a posterior distribution.
Shirai et al. considered melody generation as a sampling problem from the posterior distribution and derived Gibbs sampling and automatic composition based on it~\cite{Shirai2011}.
If we integrate constraints such as chord progressions and lyrics via PGMs, we can create an automatic composition model that considers various musical components~\cite{Shirai2013}.

Since the mid-2010s, recurrent neural networks have been actively used to model sound sequences in response to the success of deep learning.
Recurrent neural networks can easily model time-series data without paying attention to context length in n-gram models.
In particular, a variational autoencoder (VAE) is a probabilistic generative model that can use a variety of neural networks in its network architecture~\cite{jiang2020transformer,dieguez2020variational,akbari2018semi}.
This is highly compatible with the PGM-based approach described previously.
In recent years, many studies have used transformers, which have demonstrated high performance in natural language processing and image recognition~\cite{huang2018music,huang2020pop}.

Owing to the rise of deep learning, automatic composition has been gaining momentum~\cite{briot2017deep}. However, rather than deep learning itself, the essential question in automatic composition is how to model the dependencies between the transition patterns of sound sequences, chord progressions, and other musical elements. 
Since then, the idea of generating music through sampling has remained unchanged.

In recent years, language models have moved in the direction of large-scale models called large-scale language models or foundation models~\cite{brown2020language,foundation}. Thus, their capabilities have become apparent. Applications for automatic composition have also been developed. However, there is no doubt that these are only generative models of the sound sequences themselves and are in line with the framework of the above discussion.

\section{Symbol Emergence Systems and emergence of semiotic meanings}
\subsection{Symbol emergence systems}
\begin{figure*}[t]
\centering
\includegraphics[width=0.6\linewidth]{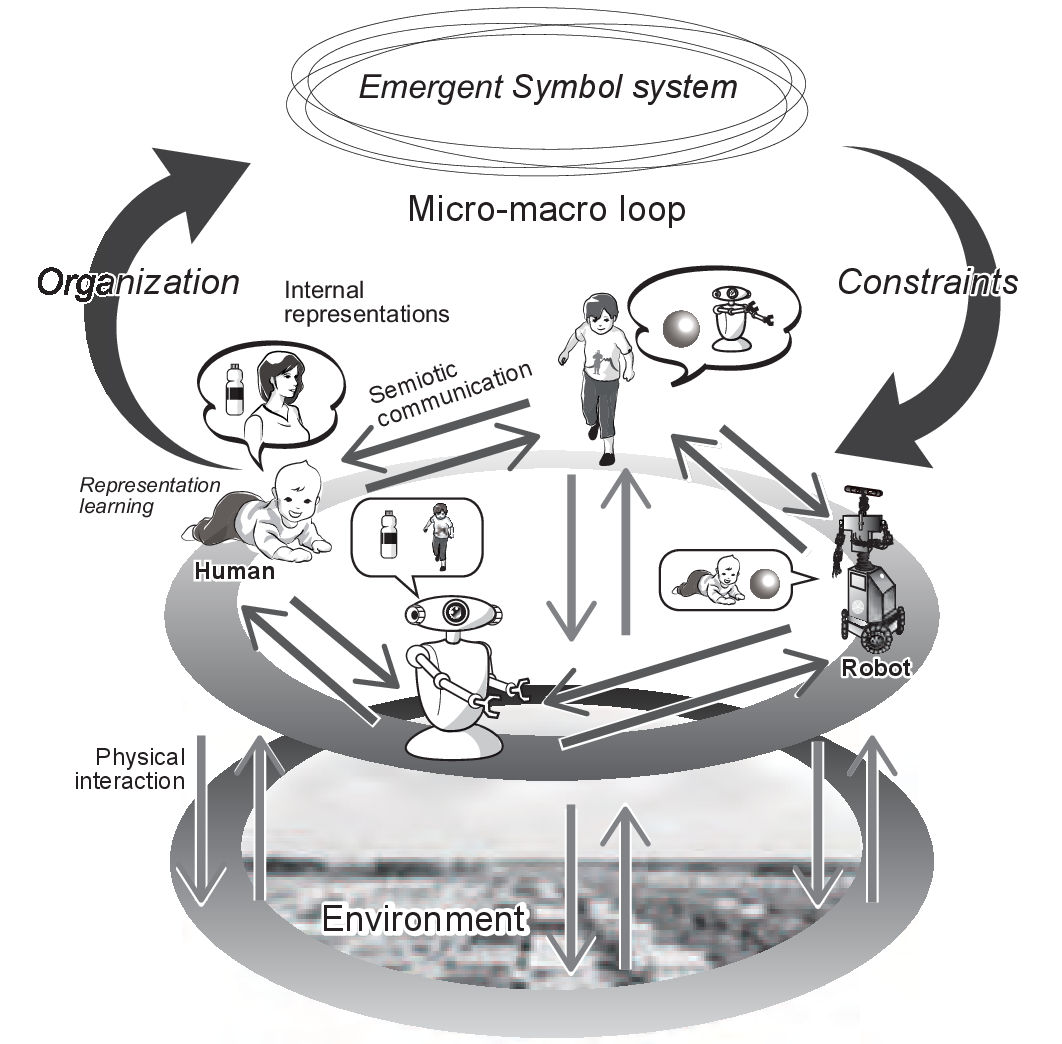}
\caption{an overview of a symbol emergence system~\cite{Taniguchi2016SER,taniguchi2018symbol}}
\label{fig:ses}
\end{figure*}
Symbol emergence systems are schematic models that describe the process through which symbols acquire meaning in society~\cite{taniguchi2018symbol}.
The symbols used are arbitrary. The relationship between a sign and an object cannot be determined a priori. When another person utters a new sign (such as a sound sequence), because we cannot look into the mind of the other person, we cannot know with certainty what the speaker means, and we can only infer.
If we accept such a reality in semiotics, how can we share the meaning of symbols in our society through an autonomous and decentralized adaptation of each agent?

A symbol emergence system is a multi-agent system consisting of multiple agents with the capability of learning generative symbols, that is, using signs for communication.
Agents form internal representations based on their interactions with their environment. In the terminology of Piaget's genetic epistemology, this corresponds to the formation of a schema~\cite{Flavell,Taniguchi2007}.
Barsalou's perceptual symbol system corresponds to the formation of symbols~\cite{Barsalou1999}. Based on the terminology of modern artificial intelligence, this can be called representation learning based on multimodal sensor information~\cite{suzuki2022survey}.
This allows the agent to form categories or concept-like objects independently. However, these internal representations are not ``symbols'' that can be used to communicate with others.
According to Peirce's definition a symbol is a triadic relationship among signs, objects, and interpretants ~\cite{Chandler2002,Peirce}. Letters in the written language and sound sequences in the spoken language are signs.
The kind of internal representation in the brain related to the sound sequence as a sign is also arbitrary.
If this can be coordinated through communication and interaction between agents, they will form a symbolic system and communicate symbolically. Consequently, an emergent symbol system was organized.

\subsection{Collective Predictive Coding}

How does an agent understand the meaning of the other's words without looking inside the other's head? Additionally, how can such a learning process lead to the emergence of stable symbol emergence systems in society?
To answer this question, the author proposes the hypothetical idea that ``symbol emergence in a society is a {\it collective predictive coding}.'' This can be regarded as a distributed Bayesian inference or social representation learning~\cite{taniguchi2022emergent}.
The author and colleagues introduced a language game called the {\it Metropolis Hastings Naming Game} and demonstrated that the emergence of word meanings and their sharing within a group can be regarded as distributed expressive learning by the entire group~\cite{taniguchi2022emergent}.
The algorithm for decentralizing the cognitive system is mathematically equivalent to the naming game, based on the idea of an integrated cognitive system that ``combines the brains'' of agents participating in communication using the same symbol system.

This suggests that the stochastic generative model is a framework that can express the ``emergence of the meaning of symbols.''
In the discussion of ''where does the meaning of a symbol come from?'', people tend to consider cognitive and social perspectives. Although the answer is apparently ``both,'' people in most scientific communities tend to focus on one of them depending on the academic community. However, the theory of symbol emergence systems emphasizes that considering both of them within one integrated model is important.

\section{On Parallelism in Music and Language}
\subsection{Parallelism on syntax, brain and evolution}

\begin{figure*}[t]
\centering
\includegraphics[width=1.0\linewidth]{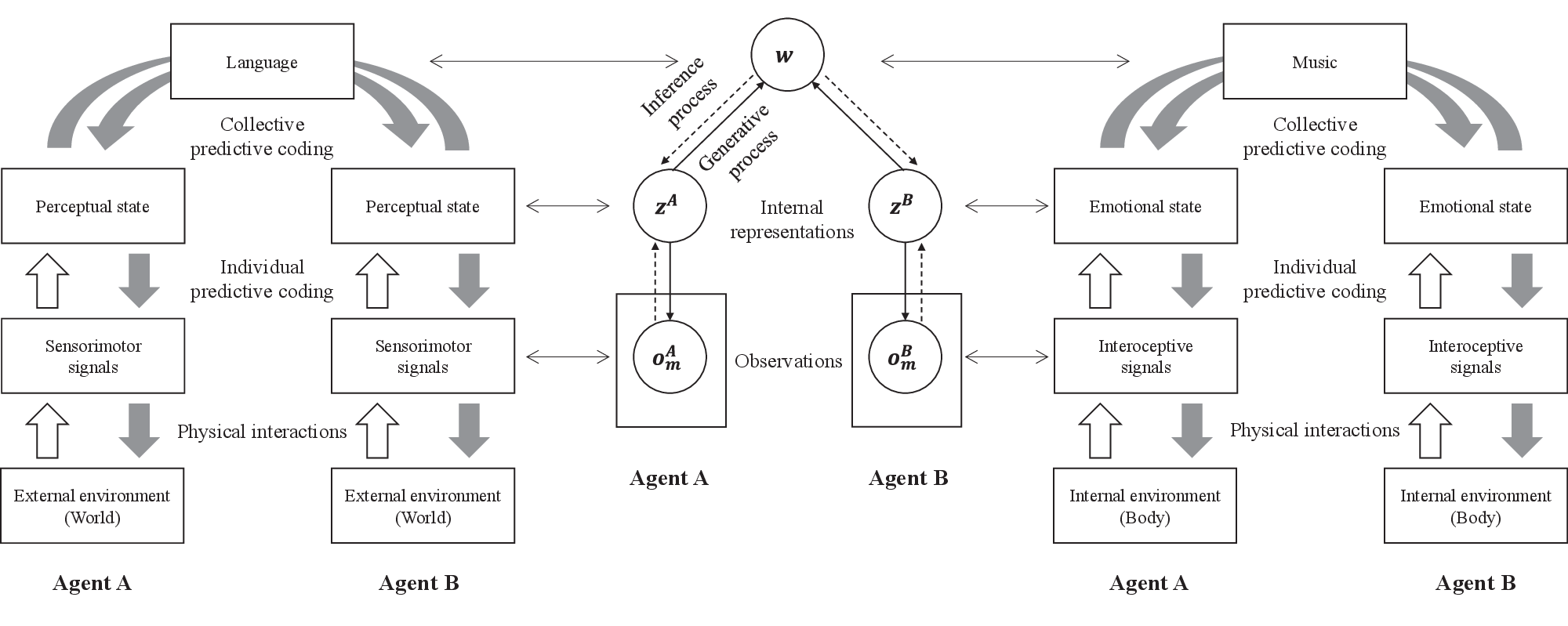}
\caption{(Left) Diagram of the process symbol emergence based on collective predictive coding. Each agent forms internal representations reflecting their perceptual state, and they use and form language through semiotic communications. The complete process is regarded as collective predictive coding~\cite{taniguchi2022emergent,Hagiwara2019,hagiwara2022multiagent}. (Center) A PGM for symbol emergence systems corresponding to Inter-DM, Inter-MDM, and Inter-GMM+VAE~\cite{Hagiwara2019,hagiwara2022multiagent,taniguchi2022emergent}. Metropolis Hastings naming game becomes a decentralized Bayesian inference of the shared $w$ and internal representations $z^A$ and $z^B$. Note that in this graphical model, head-to-head connection across $w$ is adopted~\cite{KazumaFurukawa2022}. (Right) Hypothetical diagram of the process of emergence of musical symbol systems. Instead of sensorimotor signals, emotional states are inferred through predictive coding of introspective signals~\cite{seth2013interoceptive,barrett2015interoceptive}. Such internal representations may become the basis of the emergence of a musical symbol system, i.e., the source of the socially constructed meaning of music.}
\label{fig:pgm}
\end{figure*}

It has long been argued that there is parallelism between music and language \cite{feld1994music,jackendoff1982grammatical,brown2001music}.
This relationship has been widely discussed from  anthropological, cultural, semiotic, linguistic, and musicological perspectives. This discussion is not monolithic. 

The structural similarities between music and language are often discussed, particularly with a focus on syntax~\cite{asano2015syntax}. From a musicological perspective, it has been pointed out that the grammatical structures latent in music are, in a sense, similar to those of language (although they do not have double articulation structures or phrase structure grammars, and their nature is very different). Syntactic parallelism is explained in Section 2 to a certain extent. 

In neuroscience, the similarities and differences between language and music processing have been discussed and examined~\cite{brown2006music,vuust2022music,sternin2021effect,atherton2018shared}. 

Furthermore, from the perspective of language evolution, it has been argued that an increase in vocalization complexity, that is, songs, may be a precursor to language evolution~\cite{okanoya2017sexual}. The influential hypothesis is based on evidence that birdsongs have a certain degree of grammatical structure~\cite{berwick2012bird,okanoya2007language}.

Unlike syntax, little has been said about the semantics of music in terms of its parallelism with the meaning in language~\cite{asano2015syntax}. It is difficult to discuss the semantics of music. This difficulty seems to indicate a difference between music and language in semantics, that is, meaning. However, it is difficult to discuss the meaning on the linguistic side as well. In linguistics and natural language processing, the meaning of words can be discussed almost exclusively in terms of distributional semantics or relationships with other syntactic representations, that is, semantic parsing. This means that the meaning of a symbol is considered in terms of the relationship between symbols, not in terms of their relationship with sensorimotor experiences based on interactions with the external world.

In contrast, symbol emergence systems represent integrative system dynamics in which humans form concepts based on sensory-motor systems, and meanings are determined at the social level through social interactions, that is, semiotic communications. As a constructive approach to this end, a series of studies on symbol emergence in robotics have been conducted~\cite{Taniguchi2016SER}.
This approach attempts to reproduce the emergence of symbols by addressing both the cognitive and social dynamics in the system using machine learning and robot models.
This was recently found to be related to predictive coding~\cite{hohwy2013predictive}. In addition, its relationship to the theory of self-organizational systems about neural and cognitive systems, such as the free-energy principle, has been gradually recognized~\cite{friston2021world}.

\subsection{Symbol emergence systems on music, emotion, and interoception}
The issue of the ``meaning of music'' is difficult to deal with.
Unlike syntactic structure, which can be discussed based on observable acoustic units, the meaning itself is observable. Therefore, it is difficult to reach a consensus on the definition of ``meaning'' in music.
However, ``meaning'' is also difficult to deal with in language.
If we define ``meaning'' as a dyadic relation between a sign and an object, and if we assume that the symbol system we use is static, the picture becomes relatively simple.
This view is often assumed in artificial intelligence studies such as image recognition. This may be referred to as Plato's idealistic worldview. However, such static pictures do not capture developmental language acquisition. This view cannot capture the language evolution in which the language itself emerges.
From the perspective of symbol emergence systems, the question of the meaning of language is difficult. The author believes that the viewpoint of symbol emergence systems is necessary to answer the question ``what is the meaning of language.'' If we assume the parallelism in music and language, we may be able to obtain some suggestions about the ``meaning of music'' from the perspective of symbol emergence systems.

If we have one of the most naïve viewpoints, the meaning of a word can be the speaker's internal intention or the object that the speaker means. In the model of symbol emergence systems, it is necessary for the listener to organize an internal representation system prior to communication to interpret it. Symbol emergence systems are adjusted (or emerged) through interactions between distributed agents.
They coordinate the received signs with internal representations formed through their own sensory and motor experiences.

At this point, internal representation systems are formed based on physical interactions.
The schematic representation of symbol emergence systems in Figure~\ref{fig:ses} depicts the physical interactions as interactions with the external environment. In reality, however, the ``world'' is originally assumed to be {\it Umwelt}, i.e., the subjective world and not necessarily the external world~\cite{Uexkull}. 
Our experience is not only based on the interaction with the external environment, but also on the internal environment sensed by interoception.

As mentioned earlier, it is difficult to define the ''meaning of music.” However, if we view the ''meaning of music'' as a change in the mental state that the listener undergoes, or inference (or state updating) of internal representations, similar to the ''meaning of language,'' and especially if we view it as an emotional impression (i.e., being moved, or its effect on the emotions), then through the discussion of predictive coding, we can connect it to the discussion of symbol emergence systems, which can be connected to the discussion of symbol emergence systems through the discussion of predictive coding.

In recent years, it has been argued that emotions are based on the predictive encoding of visceral sensations and introspective signals~\cite{seth2013interoceptive,barrett2015interoceptive}.
Based on this, we would like a hypothetical perspective on the correspondence between music and language, as shown in Figure~\ref{fig:pgm}.
The sign of music corresponds to the sign of language. The ''perceptual'' internal representation system that supports the interpretation of language corresponds to the ''emotional'' internal representation system in music.
Perceptual internal representational systems are organized to predict sensorimotor information caused by an external environment. In contrast, emotional internal representational systems are organized to predict sensorimotor information caused by the internal environment. 
Language can also represent emotional states because it is a highly multifaceted system. This paper shows this comparative schema to give a clearer contrast, that is, parallelism, between music and language, assuming the argument that music does not have a function explicitly representing events in the external world, unlike human language. 
In this way, the discussion of symbol emergence systems on the emergence and acquisition of language can be mapped to the emergence of symbol systems such as music.

The central figure of Figure~\ref{fig:pgm} shows the PGM representing the symbol emergence system as a whole. This view was introduced in recent studies ~\cite{Hagiwara2019,hagiwara2022multiagent,taniguchi2022emergent}.
The symbol emergence between agents A and B can be described as a decentralized Bayesian inference.
\begin{align}
 \bm{w} &\sim p(\bm{w}|\bm{z^A},\bm{z^B})p(\bm{z^A}|\{ \bm{o^A_m}\}) p(\bm{z^A}|\{ \bm{o^A_m}\})\label{eq:ses}
\end{align}
where $p(\bm{w}|\bm{z^A},\bm{z^B})$ can be sampled using a type of decentralized language game ~\cite{taniguchi2022emergent}. 

This correspondence provides an initial step in thinking about the meaning of music from the viewpoint of symbol emergence systems. This paper does not provide further details or evidence of this correspondence. However, the author believes that this perspective certainly has the potential to evoke new discussions about parallelism in music and language.

\section{Conclusion}
This article introduces generative models for symbol emergence based on real-world sensorimotor information and communication, which have been developed in a series of studies on symbol emergence in robotics. The paper also describes the symbol emergence systems that form the background of these models and the proximity of these models to models of automatic composition.
Based on the above, this paper introduced the idea that symbol emergence systems can be regarded as a multi-agent system performing collective predictive coding.
By combining this idea with the hypothesis that emotions can be explained by the predictive encoding of visceral sensory stimuli, I proposed a new view of ''the meaning of music'' as mediated by emotions and connected to symbol emergence systems.

There is one important difference between music and language from the viewpoint of symbol emergence. The sign of language has no direct influence on the external sensory systems that are directly related to the meaning of the linguistic sign. For example, the word ``apple'' does not give any direct sensation of a fruit ``apple.'' The perceptual concept of an ``apple'' is based on visual, haptic, and taste sensory information.  
In contrast, music tends to have a relatively direct influence on the internal sensory stimuli.
Peirce classified symbols as firstness, secondness, and thirdness~\cite{Peirce}. A symbol with complete arbitrariness is given meaning by the arbitrary triadic relationship between the sign and object, i.e., thirdness. In contrast, firstness means that the sign itself has a reason for ``meaning.'' For example, a sequence of sounds synchronized with the heartbeat affects the visceral senses on its own. This implies that music has many symbolic aspects of firstness. On the other hand, it is also true that music is a symbol of cultural aspects, as evidenced by the fact that musical trends change over time and that music reflects the time; for example, the ‘80s mood. 
Thus, we conclude the discussion of the parallelism of music and language by combining the viewpoints of semiotics and computational models, especially probabilistic generative models. 

In the context of artificial intelligence research, the author pointed out the confusion in the view of symbols in the symbol grounding problem and how the symbol emergence problem is a problem to be discussed~\cite{taniguchi2018symbol}. The picture of symbol emergence systems describes the process by which signs that previously had no meaning take on meaning as emergent phenomena within multi-agent systems composed of humans. The connection to the ''meaning of music'' discussed in this paper is a highly hypothetical picture that originates from outside research communities, such as music informatics and musicology, which are more closely connected to music.
However, to discuss the elusive subject of ''the meaning of music”, interdisciplinary thinking will provide suggestions. In addition, thinking about the emergence of music will be beneficial for understanding symbol emergence in a general sense, e.g., cultural and historical symbol systems.

\bibliographystyle{splncs04}
\bibliography{taniguchi}
\end{document}